\documentclass[%
 aip,
 jmp,%
 amsmath,amssymb,
 preprint,%
prl]{revtex4-1}

\usepackage{graphicx}% Include figure files
\usepackage{dcolumn}% Align table columns on decimal point
\usepackage{bm}% bold math
\begin{document}
\newcounter{subfigure}
%\preprint{AIP/123-QED}

\title[High Frequency Geodesic Acoustic Modes in Electron Temperature Gradient Mode Turbulence]{High Frequency Geodesic Acoustic Modes in Electron Temperature Gradient Mode Turbulence}% Force line breaks with \\
%\thanks{Footnote to title of article.}

\author{Johan Anderson}
\email{anderson.johan@gmail.com.}
\affiliation{ 
Department of Nuclear Engineering, Chalmers University of Technology, SE-412 96 G\"{o}teborg, Sweden
}%
\author{Hans Nordman}%
\affiliation{ 
Department of Earth and Space Sciences, Chalmers University of Technology, SE-412 96 G\"{o}teborg, Sweden
}%
\author{Raghvendra Singh} 
\affiliation{%
Institute for Plasma Research, Bhat, Gandhinagar, Gujarat, India 382428
}%
\author{Predhiman Kaw} 
\affiliation{%
Institute for Plasma Research, Bhat, Gandhinagar, Gujarat, India 382428
}%

\date{\today}% It is always \today, today,
             %  but any date may be explicitly specified

\begin{abstract}
In this work the first demonstration of a high frequency branch of the geodesic acoustic mode (GAM) driven by electron temperature gradient (ETG) modes is presented. The work is based on a fluid description of the ETG mode retaining non-adiabatic ions and the dispersion relation for high frequency GAMs driven nonlinearly by ETG modes is derived. A new saturation mechanism for ETG turbulence through the interaction with high frequency GAMs is found, resulting in a significantly enhanced ETG turbulence saturation level compared to the mixing length estimate. 
\end{abstract}

% PACS, the Physics and Astronomy
                             % Classification Scheme.
\maketitle

\section{\label{sec:level1} Introduction}
There has been overwhelming evidence that coherent structures such as vortices, streamers and zonal flows ($m=n=0$, where $m$ and $n$ are the poloidal and toroidal modenumbers respectively) play a critical role in determining the overall transport in magnetically confined plasmas.\cite{diamond2005,terry2000} Some of these coherent structures, so called streamers, are radially elongated structures that cause intermittent, bursty events, which can mediate significant transport of heat and particles, for instance, imposing a large heat load on container walls. Zonal flows on the other hand may impede transport by shear decorrelation.\cite{diamond2005,terry2000} The Geodesic Acoustic Mode (GAM)\cite{winsor1968, itoh2005, miki2007, chak2007, miki2010, hallatschek2012, mckee2008, fujisawa2007, conway2009, dong2009} is the oscillatory counterpart of the zonal flow ($m=n=0$ in the potential perturbation, $m=1$, $n=0$ in the perturbations in density, temperature and parallel velocity) and thus a much weaker effect on turbulence is expected. Nevertheless experimental studies suggest that GAMs ($n=0$, $m=1$) are related to the L-H transition and transport barriers. The GAMs are weakly damped by Landau resonances and moreover this damping effect is weaker at the edge suggesting that GAMs are more prominent in the region where transport barriers are expected. \cite{mckee2008}

The electron-temperature-gradient (ETG) mode driven by a combination of electron temperature gradients and field line curvature effects is a likely candidate for driving electron heat transport.\cite{liu1971, horton1988, jenko2000, singh2001, tangri2005, nevins2006} The ETG driven electron heat transport is determined by short scale fluctuations that do not influence ion heat transport and is largely unaffected by the large scale flows stabilizing ion-temperature-gradient (ITG) modes.

In this work the first demonstration of a high frequency branch of the geodesic acoustic mode (GAM) driven by electron temperature gradient (ETG) modes is presented. We have utilized a fluid model for the ETG mode based on the Braghinskii equations with non-adiabatic ions including impurities and finite $\beta$ - effects. \cite{singh2001, tangri2005} A new saturation mechanism for ETG turbulence through the interaction with high frequency GAMs, balanced by Landau damping, is found, resulting in a significantly enhanced  ETG turbulence saturation level compared to the mixing length estimate. 

The remainder of the paper is organized as follows: In Section II the linear ETG mode including the ion impurity dynamics is presented. The linear high frequency GAM is presented and the non-linear effects are discussed in Section III, whereas the saturation mechanism for the ETG turbulence is treated in Section IV. The paper is concluded in Section VI.

\section{\label{sec:level2} The linear Electron Temperature Gradient Mode}
In this section we will describe the preliminaries of the electron-temperature-gradient (ETG) mode which is present under the following restrictions on real frequency and wave length: $\Omega_i \leq \omega \sim \omega_{\star} << \Omega_e$, $k_{\perp} c_i > \omega > k_{||} c_e$. Here $\Omega_j$ are the respective cyclotron frequencies, $\rho_j$ the Larmor radii and $c_j = \sqrt{T_j/m_j}$ the thermal velocities. The diamagnetic frequency is $\omega_{\star} \sim k_{\theta} \rho_e c_e / L_n$, $k_{\perp}$ and $k_{||}$ are the perpendicular and the parallel wavevectors. The ETG model consists of a combination of an ion and electron fluid dynamics coupled through the quasineutrality including finite $\beta$-effects. \cite{singh2001, tangri2005}

\subsection{\label{sec:level3} Ion and impurity dynamics}
In this section, we will start by describing the ion fluid dynamics in the ETG mode description, in the limit $\omega > k_{\parallel} c_e$ the ions are stationary along the mean magnetic field $\vec{B}$ (where $\vec{B} = B_0 \hat{e}_{\parallel}$) whereas in the limit $k_{\perp} c_i >> \omega$, $k_{\perp} \rho_i >> 1$ the ions are unmagnetized. We note that the adiabatic ion response follows from the perpendicular ion momentum equation by balancing the linear parts of,
\begin{eqnarray} \label{eq:0.1}
-en_i \nabla \phi = T_i \nabla n_i,
\end{eqnarray}
and we find
\begin{eqnarray} \label{eq:0.2}
\tilde{n}_i = - \tau \tilde{\phi}.
\end{eqnarray}
In this paper we will use the non-adabatic responses in the limits $\omega < k_{\perp} c_I < k_{\perp} c_i$, $\tau_I = \sqrt{\frac{T_I}{m_I}}$ and $\Omega_i < \omega < \Omega_e$ are fulfilled for the ions and impurities. In the ETG mode description we can utilize the ion and impurity continuity and momentum equations of the form,
\begin{eqnarray} \label{eq:1.1}
\frac{\partial \tilde{n}_j}{\partial t} + n_j \nabla \cdot \vec{v}_j & = & 0, \ \ \mbox{and} \\
m_j n_j \frac{\partial v_j}{\partial t} + e n_j \nabla \phi + T_j \nabla n_j & = & 0,
\end{eqnarray}
where $j=i$ for ions and $j=I$ for impurities. Now, we derive the non-adiabatic ion response
with $\tau_i = T_e/T_i$ and impurity response with with $\tau_I = T_e/T_I$, respectively. We have thus,
\begin{eqnarray}
\tilde{n}_j = - \left( \frac{z \tau_j}{1 - \omega^2/(k_{\perp}^2 c_j^2)}\right) \tilde{\phi}, \label{eq:1.2}
\end{eqnarray}
Here $T_j$ and $n_j$ are the mean temperature and density of species ($j=e,i,I$), where $\tilde{n}_i = \delta n/ n_i$, $\tilde{n}_I = \delta n_I/ n_I$ and $\tilde{\phi} =  e \phi/T_e$ are the normalized ion density, impurity density and potential fluctuations. We also define $z_{eff} = \sum z_k \tilde{n}_k/\sum z_k \tilde{n}_k \approx (\tilde{n}_i + z^2 \tilde{n}_I)/n_e$. Next we present the electron dynamics and the linear dispersion relation.
\subsection{\label{sec:level4} The electron model}
The electron dynamics for the toroidal ETG mode are governed by the continuity, parallel momentum and energy equations adapted from the Braghinskii's fluid equations. The electron equations are analogous to the ion fluid equations used for the toroidal ITG mode,
\begin{eqnarray}
\frac{\partial n_{e}}{\partial t} +\nabla \cdot \left( n_{e} \vec{v}_{E} + n_{e} \vec{v}_{\star e} \right) + \nabla \cdot \left( n_{e} \vec{v}_{pe} + n_{e} \vec{v}_{\pi e} \right) + \nabla \cdot (n_e v_{|| e}) & = & 0 \label{eq:1.4} \\ 
 \frac{3}{2} n_{e} \frac{d T_{e}}{dt} + n_{e} T_{e} \nabla \cdot \vec{v}_{e} + \nabla \cdot \vec{q}_{e} & = & 0. \label{eq:1.5}
\end{eqnarray}
Here we used the definitions $\vec{q}_e = - (5 p_e/2m_e \Omega_e) e_{||} \times \nabla T_e$ as the diamagnetic heat flux, $v_{E}$ is the $\vec{E} \times \vec{B}$ drift, $v_{\star e}$ is the electron diamagnetic drift velocity, $\vec{v}_{Pe}$ is the polarization drift velocity, $\vec{v}_{\pi}$ is the stress tensor drift velocity, and the derivative is defined as $d/dt = \partial/\partial t + \rho_e c_e \hat{e} \times \nabla \phi \cdot \nabla$. A relation between the parallel current density and the parallel component of the vector potential ($J_{\parallel}$) can be found using Amp\`{e}re's law,
\begin{eqnarray}\label{eq:1.6}
\nabla^2_{\perp} \tilde{A}_{\parallel} = - \frac{4 \pi}{c} J_{\parallel},
\end{eqnarray}
Simplifying and linearizing Eqs. (\ref{eq:1.4}, \ref{eq:1.5} and \ref{eq:1.6}) we find the evolution equations for the electron density and the electron temperature in normalized form,
\begin{eqnarray} 
- \frac{\partial \tilde{n}_e}{\partial t}  - \nabla_{\perp}^2 \frac{\partial}{\partial t} \tilde{\phi} - \left( 1  + (1 + \eta_e) \nabla_{\perp}^2\right)\nabla_{\theta} \tilde{\phi} - \nabla_{||} \nabla_{\perp}^2 \tilde{A}_{||}  & + & \nonumber  \\ \epsilon_n \left( \cos \theta \frac{1}{r}\frac{\partial}{\partial \theta} + \sin \theta \frac{\partial}{\partial r} \right)(\tilde{\phi} - \tilde{n}_e - \tilde{T}_e) & = & 0, \label{eq:1.101} \\ 
\left((\beta_e/2 - \nabla_{\perp}^2) \frac{\partial}{\partial t} + (1+\eta_e)(\beta_e/2)\nabla_{\theta}\right)\tilde{A}_{||} + \nabla_{||} (\tilde{\phi} - \tilde{n}_e - \tilde{T}_e) & = & 0, \label{eq:1.102}\\
\frac{\partial}{\partial t}\tilde{T}_e + \frac{5}{3} \epsilon_n \left( \cos \theta \frac{1}{r}\frac{\partial}{\partial \theta} + \sin \theta \frac{\partial}{\partial r} \right) \frac{1}{r}\frac{\partial}{\partial \theta} \tilde{T}_e + (\eta_e - \frac{2}{3}) \frac{1}{r}\frac{\partial}{\partial \theta} \tilde{\phi} - \frac{2}{3} \frac{\partial}{\partial t} \tilde{n}_e & = & 0 \label{eq:1.103}. 
\end{eqnarray}
The variables are normalized according to
\begin{eqnarray}
(\tilde{\phi}, \tilde{n}, \tilde{T}_e) & = & (L_n/\rho_e)(e \delta \phi/T_{eo}, \delta n_e/n_0, \delta T_e/T_{e0}), \\ \label{eq:1.11}
\tilde{A}_{||} & = & (2 c_e L_n/\beta_e c \rho_e) e A_{||}/T_{e0}, \\ \label{eq:1.12}
\beta_e & = &  8 \pi n T_e/B_0^2.  \label{eq:1.13}
\end{eqnarray}
Using the Poisson equation in combination with (\ref{eq:1.2}) we then find
\begin{eqnarray}
\tilde{n}_e = - \left( \frac{\tau_i \tilde{n}_i/n_e}{1 - \omega^2/k_{\perp}^2 c_i^2} + \frac{(Z^2\tilde{n}_I/n_e) \tau_I}{1 - \omega^2/(k_{\perp}^2 c_I^2)} + k_{\perp}^2 \lambda_{De}^2\right) \tilde{\phi}. \label{eq:1.3}
\end{eqnarray}
First we will consider the linear dynamical equations (\ref{eq:1.101}, \ref{eq:1.102} and \ref{eq:1.103}) and utilizing Eq. (\ref{eq:1.5}) in the same manner as in Refs \onlinecite{singh2001, tangri2005} and we find a semi-local dispersion relation as follows,
\begin{eqnarray} \label{eq:1.14}
\left[ \omega^2 \left(  \Lambda_e  + \frac{\beta_e }{2} (1 + \Lambda_e ) \right) + \left( 1 - \bar{\epsilon}_n (1 + \Lambda_e) \right) \omega_{\star} \right. & + & \nonumber \\
\left. k_{\perp}^2 \rho_e^2 \left(  \omega - (1 + \eta_e) \omega_{\star} \right) \right] \left( \omega - \frac{5}{3} \bar{\epsilon}_n \omega_{\star} \right) & + & \nonumber \\
\left( \bar{\epsilon}_n \omega_{\star} - \frac{\beta_e}{2} \omega\right) \left( (\eta_e - \frac{2}{3})\omega_{\star} + \frac{2}{3} \omega \Lambda_e \right) & = & \nonumber \\
c_e^2k_{||}^2 k_{\perp}^2 \rho_e^2 \left( \frac{(1 + \Lambda_e) \left( \omega - \frac{5}{3}\bar{\epsilon}_n \omega_{\star}\right) - \left( \eta_e - \frac{2}{3}\right)\omega_{\star} - \frac{2}{3} \omega \Lambda_e}{\omega \left( \frac{\beta_e}{2} + k_{\perp}^2 \rho_e^2\right) - \frac{\beta_e}{2} (1 + \eta_e) \omega_{\star}} \right) 
\end{eqnarray}
In the following we will use the notation $\Lambda_e = \tau_i (n_i/n_e)/(1 - \omega^2/k_{\perp}^2 c_i^2) + \tau_I (Z_{eff} n_I/n_e)/(1 - \omega^2/k_{\perp}^2 c_I^2) + k_{\perp}^2 \lambda_{De}^2$. Note that in the limit $T_i = T_e$, $\omega<k_{\perp} c_i$, $k_{\perp} \lambda_{De} < k_{\perp} \rho_e \leq 1$ and in the absence of impurity ions, $\Lambda_e \approx 1$ and the ions follow the Boltzmann relation in the standard ETG mode dynamics. Here $\lambda_{De} = \sqrt{T_c/(4 \pi n_e e^2)}$ is the Debye length, the Debye shielding effect is important for $\lambda_{De}/\rho_e > 1$ \cite{singh2001}. The dispersion relation Eq. (\ref{eq:1.14}) is analogous to the toroidal ion-temperature-gradient mode dispersion relation ecept that the ion quantities are exchanged to their electron counterparts. In Eq. (\ref{eq:1.14}), the geometrical quantities will be determined using a semi-local analysis by assuming an approximate eigenfunction while averageing the geometry dependent quantities along the field line. The form of the eigenfunction is assumed to be\cite{jarmen1998},
\begin{eqnarray}\label{eq:1.15}
\Psi(\theta) = \frac{1}{\sqrt{3 \pi}}(1 + \cos \theta) \;\;\;\;\; \mbox{with} \;\;\;\;\; |\theta| < \pi.
\end{eqnarray}
In the dispersion relation we will replace $k_{\parallel} = \langle k_{\parallel} \rangle$, $k_{\perp} = \langle k_{\perp} \rangle$ and $\omega_D = \langle \omega_D \rangle$ by the averages defined through the integrals, 
\begin{eqnarray} 
\langle k_{\perp}^2 \rangle & = & \frac{1}{N(\Psi)}\int_{-\pi}^{\pi} d \theta \Psi k_{\perp}^2 \Psi = k_{\theta}^2 \left( 1 + \frac{s^2}{3} (\pi^2 - 7.5) - \frac{10}{9} s \alpha + \frac{5}{12} \alpha^2 \right), \label{eq:1.16} \\
\langle k_{||}^2 \rangle & = & \frac{1}{N(\Psi)} \int_{-\pi}^{\pi} d \theta \Psi k_{||}^2 \Psi = \frac{1}{3 q^2 R^2}, \label{eq:1.17}\\
\langle \omega_D \rangle & = & \frac{1}{N(\Psi)} \int_{-\pi}^{\pi} d \theta \Psi \omega_D \Psi = \epsilon_n \omega_{\star} \left( \frac{2}{3} + \frac{5}{9}s - \frac{5}{12} \alpha \right),  \label{eq:1.17} \\
\langle k_{\parallel} k_{\perp}^2 k_{\parallel} \rangle & = & \frac{1}{N(\Psi)} \int_{-\pi}^{\pi} d \theta \Psi k_{\parallel} k_{\perp}^2 k_{\parallel} \Psi = \frac{k_{\theta}^2}{3 (qR)^2} \left( 1 + s^2 (\frac{\pi^2}{3} - 0.5) - \frac{8}{3}s \alpha + \frac{3}{4} \alpha^2 \right). \label{eq:1.18} \\
N(\Psi) & = & \int_{-\pi}^{\pi} d \theta \Psi^2.
\end{eqnarray}
Here $\alpha = \beta q^2 R \left(1 + \eta_e + (1 + \eta_i) \right)/(2 L_n)$ and $\beta = 8 \pi n_o (T_e + T_i)/B^2$ is the plasma $\beta$, $q$ is the safety factor and $s = r q^{\prime}/q$ is the magnetic shear. The $\alpha$-dependent term above (in Eq. \ref{eq:1.16}) represents the effects of Shafranov shift. 

\section{\label{sec:level6} Modeling High Frequency Geodesic Acoustic modes}
The Geodesic Acoustic Modes are the $m=n=0$, $k_r \neq 0$ perturbation of the potential field and the $n=0$, $m=1$, $k_r \neq 0$ perturbation in the density, temperatures and the magnetic field perturbations. The high frequency GAM ($q, \Omega$) induced by ETG modes ($k,\omega$) is prevailing under the conditions when the ETG mode real frequence satisfies ($\Omega_e > \omega > \Omega_i$) at the scale ($k_{\perp } \rho_i < 1$) and the real frequence of the GAM fulfills ($\Omega_e > \Omega > \Omega_i$) at the scale ($q_x < k_{x}$). 

\subsection{Linear Geodesic Acoustic Modes}
We start by deriving the linear electron GAM dispersion relation, by writing the $m=1$ equations for the density, parallel component of the vector potential, temperature and the $m=0$ of the electrostatic potential
\begin{eqnarray} 
- \tau_i \frac{\partial \tilde{n}^{(1)}_{eG}}{\partial t} + \epsilon_n \sin \theta \frac{\partial}{\partial r} \tilde{\phi}^{(0)}_G - \nabla_{||} \nabla_{\perp}^2 \tilde{A}_{|| G}^{(1)} & = & 0, \label{eq:3.101} \\ 
\left((\beta_e/2 - \nabla_{\perp}^2) \frac{\partial}{\partial t} + (1+\eta_e)(\beta_e/2)\nabla_{\theta}\right)\tilde{A}^{(1)}_{|| G} - \nabla_{||} (\tilde{n}^{(1)}_{eG} + \tilde{T}^{(1)}_{eG}) & = & 0, \label{eq:3.102} \\
\frac{\partial}{\partial t}\tilde{T}^{(1)}_{eG} - \frac{2}{3} \frac{\partial}{\partial t} \tilde{n}^{(1)}_{eG} & = & 0, \label{eq:3.103} \\
- \nabla_{\perp}^2 \frac{\partial}{\partial t} \tilde{\phi}^{(0)}_G - \epsilon_n \sin \theta \frac{\partial}{\partial r} (\tilde{n}^{(1)}_{eG} + \tilde{T}^{(1)}_{eG}) & = & 0. \label{eq:3.104}
\end{eqnarray} 
First we will derive the linear GAM frequency assuming electrostatic GAMs ($\beta_e \rightarrow 0$) this yields a relation between the parallel component of the vector potential and the density and electron perturbations using Eq. (\ref{eq:3.102}) as,  
\begin{eqnarray} \label{eq:3.3}
- \nabla_{\perp}^2 \frac{\partial \tilde{A}_{|| G}^{(1)}}{\partial t} - \nabla_{||} (\tilde{n}_G^{(1)} + \tilde{T}_{e G}^{(0)}) = 0.  
\end{eqnarray}
The $m=1$ component of the electron density can be eliminated by taking a time derivative of Eq. (\ref{eq:3.104}) and using Eq. (\ref{eq:3.101}) and we get,
\begin{eqnarray} \label{eq:3.8}
\rho_e^2 \frac{\partial^2}{\partial t^2} \nabla_{\perp}^2 \tilde{\phi}^{(0)} +  \epsilon_n v_{\star} \langle \sin \theta \frac{\partial}{\partial r} \left( \epsilon_n v_{\star} \sin \theta \frac{\partial \tilde{\phi}^{(0)}}{\partial r} + \nabla_{||} \frac{J_{||}^{(1)}}{e n_0} \right) \rangle = 0.
\end{eqnarray}
Here $\langle \cdots \rangle$ is the average over the poloidal angle $\theta$. In the simplest case this leads to the dispersion relation,
\begin{eqnarray}  \label{eq:3.9}
\Omega^2 = \frac{c_e^2}{R^2} \left( \frac{10}{3} + \frac{1}{q^2}\right). 
\end{eqnarray}
Note that the linear electron GAM is purely oscillating analogously to its ion counterpart. In previous section we computed the linear dispersion relation for the GAM now we will study the non-linear contributions through a modulational instability analysis.

\subsection{\label{sec:level7} The Non-linearly Driven Geodesic Acoustic modes}
We will now study the system including the non-linear terms and derive the electron GAM growth rate. The non-linear extension to the evolution equations presented previously in Eqs (\ref{eq:1.101}) - (\ref{eq:1.103}) are
\begin{eqnarray} 
- \frac{\partial \tilde{n}_e}{\partial t}  - \nabla_{\perp}^2 \frac{\partial}{\partial t} \tilde{\phi} - \left( 1  + (1 + \eta_e) \nabla_{\perp}^2\right)\nabla_y \tilde{\phi} - \nabla_{||} \nabla_{\perp}^2 \tilde{A}_{||}  & + & \nonumber  \\ \epsilon_n \left( \cos \theta \frac{1}{r}\frac{\partial}{\partial \theta} + \sin \theta \frac{\partial}{\partial r} \right)(\tilde{\phi} - \tilde{n}_e - \tilde{T}_e) & = & \nonumber \\  
-(\beta_e/2) [\tilde{A}_{||},\nabla_{||}^2 \tilde{A}_{||}] + [ \tilde{\phi}, \nabla^2 \tilde{\phi}], \label{eq:7.101}\\ 
\left((\beta_e/2 - \nabla_{\perp}^2) \frac{\partial}{\partial t} + (1+\eta_e)(\beta_e/2)\nabla_y\right)\tilde{A}_{||} + \nabla_{||} (\tilde{\phi} - \tilde{n}_e - ) & = & - (\beta_e/2) [\tilde{\phi} - \tilde{n}_e, \tilde{A}_{||}] \nonumber \\
+ (\beta_e/2)[\tilde{T}_e,\tilde{A}_{||}] + [\tilde{\phi}, \nabla_{\perp}^2 \tilde{A}_{||}], \label{eq:7.102}\\
\frac{\partial}{\partial t}\tilde{T}_e + \frac{5}{3} \epsilon_n \left( \cos \theta \frac{1}{r}\frac{\partial}{\partial \theta} + \sin \theta \frac{\partial}{\partial r} \right) \frac{1}{r}\frac{\partial}{\partial \theta} \tilde{T}_e + (\eta_e - \frac{2}{3}) \frac{1}{r}\frac{\partial}{\partial \theta} \tilde{\phi} - \frac{2}{3} \frac{\partial}{\partial t} \tilde{n}_e & = & -[\tilde{\phi},\tilde{T}_e]. \label{eq:7.103}
\end{eqnarray}
In order to find the relevant equations for the electron GAM dynamics we consider the $m=1$ component of Eqs (\ref{eq:7.101}) - (\ref{eq:7.103}) and we find,
\begin{eqnarray} \label{eq:3.1}
- \tau_i \frac{\partial \tilde{n}^{(1)}_{eG}}{\partial t} + \epsilon_n \sin \theta \frac{\partial}{\partial r} \tilde{\phi}^{(0)}_G - \nabla_{||} \nabla_{\perp}^2 \tilde{A}_{|| G}^{(1)} =  \langle [\tilde{\phi}_k, \nabla^2 \tilde{\phi}_k] \rangle^{(1)} & - &(\beta_e/2) \langle  [\tilde{A}_{|| k},\nabla_{||}^2 \tilde{A}_{|| k}]\rangle^{(1)}, \nonumber \\ \\ 
\left((\beta_e/2 - \nabla_{\perp}^2) \frac{\partial}{\partial t} + (1+\eta_e)(\beta_e/2)\nabla_{\theta}\right)\tilde{A}^{(1)}_{|| G} - \nabla_{||} (\tilde{n}^{(1)}_{eG} + \tilde{T}^{(1)}_{eG}) & = & - (\beta_e/2) \langle [\tilde{\phi}_k - \tilde{n}_{e k}, \tilde{A}_{|| k}] \rangle^{(1)} \nonumber \\
+ (\beta_e/2) \langle [\tilde{T}_{e k},\tilde{A}_{|| k}] \rangle^{(1)} + \langle [\tilde{\phi}_k, \nabla_{\perp}^2 \tilde{A}_{|| k}] \rangle^{(1)}, \\
\frac{\partial}{\partial t}\tilde{T}^{(1)}_{eG} - \frac{2}{3} \frac{\partial}{\partial t} \tilde{n}^{(1)}_{eG} & = & -\langle [\tilde{\phi}_k,\tilde{T}_{ek}]\rangle^{(1)}, 
\end{eqnarray}
where superscript (1) over the fluctuating quantities denotes the $m=1$ poloidal mode number and $\langle \cdots \rangle $ is the average over the fast time and spatial scale of the ETG turbulence and that non-linear terms associated with parallel dynamics are small since $\frac{1}{q^2} << 1$. We now study the $m=0$ potential perturbations,
\begin{eqnarray} \label{eq:3.2}
- \nabla_{\perp}^2 \frac{\partial}{\partial t} \tilde{\phi}^{(0)}_G - \epsilon_n \sin \theta \frac{\partial}{\partial r} (\tilde{n}^{(1)}_{eG} + \tilde{T}^{(1)}_{eG}) =  \langle [\tilde{\phi}_k, \nabla^2 \tilde{\phi}_k] \rangle^{(0)} & - & (\beta_e/2) \langle [\tilde{A}_{|| k},\nabla_{||}^2 \tilde{A}_{||} k] \rangle^{0}.
\end{eqnarray}
We will now neglect the effects of Debye shielding and make use of quasi-neutrality in the plasma of the form ($\tilde{n}_i = \tilde{n}_e = \tilde{n}$) and subtracting Eq. (\ref{eq:1.101}) from Eq. (\ref{eq:1.102}) we find,
\begin{eqnarray} \label{eq:3.5}
\rho_e^2 \frac{\partial}{\partial t} \nabla_{\perp}^2 \tilde{\phi} + \frac{v_{\star}}{r} \frac{\partial \tilde{\phi}}{\partial \theta} - \epsilon_n v_{\star} \left( \cos \theta \frac{1}{r} \frac{\partial}{\partial \theta} + \sin \theta \frac{\partial}{\partial r}\right) (\tilde{\phi} - \tilde{n}_e  - \tilde{T}_e) - \nabla_{||} \tilde{J}_{||} = N_1.
\end{eqnarray}
Here we have defined the non-linear term on the RHS as $N_1 =  \rho_e^3 c_e \hat{z} \times \nabla \tilde{\phi} \cdot \nabla \nabla_{\perp}^2 \tilde{\phi} + \frac{\delta B_r}{B} \cdot \nabla \frac{\tilde{J}_{||}}{e n_0}$. For the GAM we find the ($n=0, m=1$) component of Eq. (\ref{eq:1.101}) as,
\begin{eqnarray} \label{eq:3.6}
\frac{\partial \tilde{T}^{(1)}_e}{\partial t} - \frac{2}{3} \frac{\partial \tilde{n}^{(1)}_e}{\partial t} = N^{(1)}_2.
\end{eqnarray} 
This can be written $\tilde{T}_e = \frac{2}{3} \tilde{n}_e^{(1)} + N_2^{1}$, where the $m=1$ component is determined by an integral of the convective non-linear term as $N_2^{1} = - \int dt \rho_s c_s \hat{z} \times \nabla \tilde{\phi}^{(0)} \cdot \nabla \tilde{T}^{(1)}_e$. This leads to a relation between the $m=1$ component of the density and temperature fluctuations modified by a non-linear term. We continue by taking the $m=1$ component of Eq. (\ref{eq:1.102}) and $m=0$ of Eq. (\ref{eq:1.103}),
\begin{eqnarray} 
\frac{\partial \tilde{n}^{(1)}_e}{\partial t} - \frac{\nabla_{||} \tilde{J}_{||}^{(1)}}{e n_0} - \epsilon_n v_{\star} \sin \theta \frac{\partial \tilde{\phi}^{(0)}}{\partial r} = N_1^{(1)}, \label{eq:3.71}\\
\rho_e^2 \frac{\partial}{\partial t} \nabla_{\perp}^2 \tilde{\phi}^{(0)} +  \epsilon_n v_{\star} \langle \sin \theta \frac{\partial}{\partial r} \left( \frac{5}{3} \tilde{n}_e^{(1)} + N_2^{1}\right) \rangle = N_1^{(0)}. \label{eq:3.72}
\end{eqnarray}
Similar to the operations performed to find the linear electron GAM frequency we eliminate the $m=1$ component of the electron density by taking a time derivative of Eq. (\ref{eq:3.72}) this yields,
\begin{eqnarray} \label{eq:3.8}
\rho_e^2 \frac{\partial^2}{\partial t^2} \nabla_{\perp}^2 \tilde{\phi}^{(0)} +  \epsilon_n v_{\star} \langle \sin \theta \frac{\partial}{\partial r} \left( \epsilon_n v_{\star} \sin \theta \frac{\partial \tilde{\phi}^{(0)}}{\partial r} + \nabla_{||} \frac{J_{||}^{(1)}}{e n_0} + N_2^{(1)} + \frac{\partial}{\partial t} N_2^{(1)}\right) \rangle = \frac{\partial}{\partial t}N_1^{(0)}.
\end{eqnarray}
Note that this will be modified by the effects of the parallel current density ($\tilde{J}_{||}$) and the non-linear terms, however we see by inspection that on average the term $N_2^{(1)}$ does not contribute whereas the $N_1^{(0)}$ non-linearity may drive the GAM unstable. 

We will use the wave kinetic equation\cite{diamond2005, chak2007, smol2000, smol2002, krommes2000, anderson2002, anderson2006, anderson2007, hallatschek2012} to describe the background short scale ETG turbulence for $(\Omega, \vec{q}) < (\omega, \vec{k})$, where the action density $N_k = E_k/|\omega_r| \approx \epsilon_0 |\phi_k|^2/\omega_r$. Here $\epsilon_0 |\phi_k|^2$, is the total energy in the ETG mode with mode number $k$ where $\epsilon_0 = \tau + k_{\perp}^2 + \frac{\eta_e^2 k_\theta^2}{|\omega|^2}$. In describing the large scale plasma flow dynamics it is assumed that there is a sufficient spectral gap between the small scale ETG turbulent fluctuations and the large scale GAM flow. The electrostatic potential is represented as a sum of fluctuating and mean quantities
\begin{eqnarray} \label{eq:4.1}
\phi(X,x,T,t) = \Phi(X,T) + \tilde{\phi}(x,t)
\end{eqnarray}
where $\Phi(X,T)$ is the mean flow potential. The coordinates $\left( X, T\right)$, $\left( x,t \right)$ are the spatial and time coordinates for the mean flows and small scale fluctuations, respectively. The wave kinetic equation can be written as,
\begin{eqnarray} \label{eq:4.2}
\frac{\partial }{\partial t} N_k(x,t) & + & \frac{\partial }{\partial k_x} \left( \omega_k + \vec{k} \cdot \vec{v}_g \right)\frac{\partial N_k(x,t)}{\partial x} - \frac{\partial }{\partial x} \left( \vec{k} \cdot\vec{v}_g\right) \frac{\partial N_k(x,t)}{\partial k_x} \nonumber \\
& = &  \gamma_k N_k(x,t) - \Delta\omega N_k(x,t)^2.
\end{eqnarray}
We will solve Equation (\ref{eq:4.2}) by assuming a small perturbation ($\delta N_k$) driven by a slow variation for the GAM compared to the mean ($N_{k0}$) such that $N_k = N_{k 0} + \delta N_k$. The relevant non-linear terms can be approximated in the following form\cite{singh2001},
\begin{eqnarray} 
\langle [\tilde{\phi}_k, \nabla_{\perp}^2 \tilde{\phi}_k] \rangle & \approx &  q_x^2 \sum_k k_x k_y \frac{|\omega_r|}{\epsilon_0} \delta N_k(\vec{q},\Omega), \\ \ \label{eq:4.3}
\langle [ \tilde{A}_{|| k}, \nabla_{\perp}^2 \tilde{A}_{|| k}] \rangle & \approx &  q_x^2 \sum_k k_x k_y \lambda_0 \frac{|\omega_r|}{\epsilon_0} \delta N_k(\vec{q},\Omega), \\ \  \label{eq:4.4}
\langle [\tilde{\phi}_k, \tilde{T}_{e k}] \rangle & \approx & -i q_x \eta_e \sum_k \frac{k_y}{\gamma_k} \frac{|\omega_r|}{\epsilon_0} \delta N_k(\vec{q},\Omega).  \label{eq:4.5}
\end{eqnarray}
For all GAMs we have $q_x > q_y$, with the following relation between $\delta N_k$ and $\partial N_{k 0} / \partial k_x$,
\begin{eqnarray} \label{eq:4.6}
\delta N_k = -i q_x^2 k_y \phi^0_q R \frac{\partial N_{0k}}{\partial k_x} + \frac{k_y q_x T_q^{(1)} N_{0k}}{\tau_i \sqrt{(\eta_e - \eta_{eth})}}, 
\end{eqnarray}
where we have used $\delta \omega_q = k\cdot v_{Eq} \approx i (k_y q_x - k_x q_y) \phi^0_q$ in the wave kinetic equation  and the definition $R = \frac{1}{\Omega_q - q_x v_{gx} + i \gamma_k}$. Using the results from the wave-kinetic treatment we can compute the non-linear contributions to be of the form,
\begin{eqnarray}
\langle \phi, \nabla_{\perp}^2 \phi \rangle & = & -i q_x^4 \sum k_x k_y^2 \frac{|\omega_r|}{\epsilon_0} R \frac{\partial N_k}{\partial k_x} \phi_q^{(0)} + \frac{2}{3} q_x^3 \sum k_x k_y \frac{|\omega_r|}{\epsilon_0} \frac{R N_0}{\tau (\eta_e - \eta_{th e})^{1/2}} n_q^{(1)}, \\  \label{eq:4.6}
\langle \phi, T_e \rangle & = & q_x^3 \sum k_y^2 \frac{\eta_e \gamma }{|\omega_r |^2} \frac{|\omega_r|}{\epsilon_0} R \frac{ \partial N_0}{\partial k_x} \phi_q^{(0)} + i q_x^2 \sum \frac{2}{3} \frac{k_y^3 \eta_e \gamma }{|\omega_r|^2} \frac{|\omega_r|}{\epsilon_0} \frac{R N_0}{\tau (\eta_e - \eta_{th e})} n_q^{(1)}. \label{eq:4.7}
\end{eqnarray}
In order to find the non-linear growth rate of the electron GAM we need to find relations between the variables $n_G^{(1)}$, $T_G^{(1)}$ and $\phi_G^{(0)}$,
\begin{eqnarray}
n_G^{(1)} & = & - \frac{\epsilon_n q_x \sin \theta }{\Omega - q_{\parallel}^2/\Omega} \phi_G^{0}, \label{eq:4.8} \\  
T_G^{(1)} & = &  \frac{2}{3} n_G^{(1)} - \frac{2}{3} q_x^2\sum \frac{k_y^3 \eta_e \gamma}{|\omega|^2} \frac{|\omega_r|}{\epsilon_0} \frac{R N_0}{\tau (\eta_e - \eta_{th e})^{1/2}} n_q^{(1)}.  \label{eq:4.9}
\end{eqnarray}
Using Eqs. (\ref{eq:4.8}) and (\ref{eq:4.9}) in the Fourier representation of Eq. (\ref{eq:3.8}) resulting in
\begin{eqnarray} \label{eq:4.10}
\Omega q_x^2 \phi_G^{(0)} + \epsilon_n q_x \sin \theta (n_G^{(1)} + T_G^{(1)}) = i \langle \phi, \nabla_{\perp}^2 \phi \rangle^{(0)},
\end{eqnarray}
and we finally find
\begin{eqnarray} \label{eq:4.11}
\Omega^2 - k_{\parallel}^2 - \frac{5}{6} \epsilon_n^2 & = & - \frac{1}{3} \epsilon_n^2 q_x^2 \sum \frac{k_y^3 \eta_e \gamma}{|\omega|} \frac{|\omega_r|}{\epsilon_0}  \frac{R N_0}{\tau (\eta-\eta_{th e})^{1/2}} \nonumber 
\\ & + & \left( \frac{\Omega^2 - q_{\parallel}^2}{\Omega} \right) q_x^2 \sum k_x k_y^2 \frac{|\omega_r|}{\epsilon_0} R \frac{\partial N_k}{\partial k_x}.
\end{eqnarray}
Eq. (\ref{eq:4.11}) is the sought dispersion relation for the electron GAM and we solve it pertubatively by assuming $\Omega = \Omega_0 + \Omega_1$ where $\Omega_0$ is the solution to the linear part c.f. Eq. (\ref{eq:3.9}). Now we find the perturbation $\Omega_1 = i \gamma_q$ which will determine the growth rate of the GAM as,
\begin{eqnarray}
\frac{\gamma_q}{c_e/L_n} & = & i \frac{\epsilon_n^2}{6 \Omega_0} q_x^2 \rho_e^2 \sum \frac{k_y^3 \rho_e^3 \eta_e \gamma}{|\omega|} \frac{|\omega_r|}{\epsilon_0} \frac{N_0}{i \gamma} \frac{1}{(\eta_e - \eta_{th e})^{1/2}} - i \frac{5}{12} \frac{\epsilon_n^2}{\Omega_0^2} q_x^2 \rho_e^2 \sum k_x k_y^2 \rho_e^3  \frac{|\omega_r|}{\epsilon_0} \frac{1}{i \gamma} \left| \frac{\partial N_0}{\partial k_x} \right| \nonumber \\
& \approx & \frac{5}{12} \frac{q_x^2 \rho_e^2 k_y \rho_e}{\sqrt{\epsilon_n} (\eta_e - \eta_{th e})^{1/2}} \left| \phi_k^2 \right|. \label{eq:4.14}
\end{eqnarray}
In the last expression we have assumed that the GAM frequency ($\Omega_0$) can be approximated by $\Omega_0 \approx 2 c_e/R$, i.e. the linear GAM is purely oscillatory as found in Eq. (\ref{eq:3.9}). The non-linearly driven electron GAM is unstable with a growth rate depending on the saturation level $\left| \phi_k^2 \right|$ of the ETG mode turbulence. 

\section{\label{sec:level9} Saturation mechanism}
In this section we will estimate a new saturation level for the ETG turbulent electrostatic potential ($\phi_k$) by using the Landau damping in competition with the non-linear growth rate of the GAM in a constant background of ETG mode turbulence, according to the well known predator-prey models used\cite{malkov2001}, c.f. Eq. (4) in Ref (\onlinecite{miki2007}) and as well as Ref. (\onlinecite{miki2010}), 
\begin{eqnarray} \label{eq:5.1}
\frac{\partial N_k}{\partial t} & = & \gamma_k N_k - \Delta \omega N_k^2 - \gamma_1 U_G N_k \\
\frac{\partial U_G}{\partial t} & = & \gamma_q U_G - \gamma_L U_G.
\end{eqnarray}
Here we have represented the ETG mode turbulence as $N_k = |\phi_k|^2 \frac{L_n^2}{\rho_e^2}$ and $U_G = \langle \frac{ e \phi_G^{(0)}}{T_e} \frac{L_n}{\rho_e} \sin \theta \rangle$ with the following parameters $\gamma$ is the ETG mode growth rate, $\gamma_{NL}$ is the non-linear damping and $\gamma_1$ is the coupling between the ETG mode and the GAM. The Landau damping rate $\left( \gamma_L = \frac{4 \sqrt{2}}{3 \sqrt{\pi}} \frac{c_e}{qR} \right)$ is assumed to be balanced by GAM growth rate Eq. (\ref{eq:4.14}) in stationary state $ \frac{\partial N}{\partial t} \rightarrow 0$ and $ \frac{\partial U_G}{\partial t} \rightarrow 0$. In steady state find the saturation level for the ETG turbulent intensity as ($\gamma_q = \gamma_L$),
\begin{eqnarray} \label{eq:5.2}
\left|\frac{e \phi_k}{T_e} \frac{L_n}{\rho_e} \right|^2 = \frac{48 \sqrt{2}}{3 \sqrt{\pi}} \left( \frac{L_n}{q R} \right) \left( \frac{L_n}{\rho_e} \right)^2 \frac{\sqrt{\epsilon_n (\eta_e - \eta_{th e})}}{(q_x \rho_e)^2 (k_y \rho_e)}.
\end{eqnarray}
Note that this saturation level is significantly enhanced compared to the mixing length estimate,
\begin{eqnarray} \label{eq:5.3}
\left|\frac{e \phi_k}{T_e} \frac{L_n}{\rho_e} \right| \sim 10.
\end{eqnarray}
Here in this estimation we have used $L_n = 0.05$, $q = 3.0$, $R = 4$, $\epsilon_n = 0.025$, $q_x \rho_e = k_y \rho_e = 0.3$ and $\eta_e - \eta_{e th} \sim 1$. Note that the result found using a mixing length estimate is $\left| \frac{e \phi}{T_e} \frac{L_n}{\rho_e}\right| \sim 1$ significantly smaller. 

\section{\label{sec:level10} Conclusion}
In this paper we have presented the first derivation of a high frequency branch of the Geodesic Acoustic Mode (GAM). The linear dispersion relation of the high frequency GAM showed that the new branch is purely oscillatory with a frequency $\Omega \sim \frac{c_e}{R}$. To estimate the GAM growth rate, a non-linear treatment based on the wave-kinetic approach was applied. The resulting non-linear dispersion relation showed that the high frequency GAM is excited in the presence of ETG modes with a growth rate depending on the fluctuation level of the ETG mode turbulence. An analytical expression for the resulting GAM growth rate was obtained. To estimate the ETG mode fluctuation level and GAM growth, a predator-prey model was used to describe the coupling between the GAMs and small scale ETG turbulence. The stationary point of the coupled system implies that the ETG turbulent saturation level $\phi_k$ can be drastically enhanced by a new saturation mechanism, stemming from a balance between the Landau damping and the GAM growth rate. This may result in highly elevated particle and electron heat transport, relevant for the edge pedestal region of H-mode plasmas.

The present work was based on a fluid description of ETG mode turbulence, including finite beta electromagnetic effects and retaining non-adiabatic ions. A more accurate treatment based on quasi-linear and nonlinear gyrokinetic simulations is left for future work.

%\section*{References}

\end{document}